\documentclass[aps,preprint,nofootinbib]{revtex4}%
\usepackage{amsfonts}
\usepackage[utf8]{inputenc}
\usepackage{amsmath}
\usepackage{amssymb}
\usepackage{float}
\usepackage{graphicx}%
\providecommand{\U}[1]{\protect\rule{.1in}{.1in}}

\begin{document}
\title{Fully resonant scalars on asymptotically AdS wormholes}
\author{Andr\'{e}s Anabalon$^{1}$, Julio Oliva$^{2}$, Constanza Quijada$^{2}$}
\affiliation{$^{1}$Universidad Adolfo Iba\~{n}ez, Dep. de Ciencias, Facultad de Artes
Liberales, Vi\~{n}a del Mar, \ Chile.}
\affiliation{$^{2}$Departamento de F\'{\i}sica, Universidad de Concepci\'{o}n, Casilla
160-C, Concepci\'{o}n, Chile}

\begin{abstract}
In this work we show the existence of asymptotically AdS
wormhole geometries where the scalar probe has an equispaced, fully resonant
spectrum, as that of a scalar on AdS spacetime, and explore its dynamics when non-linearities are included. The spacetime is a solution of
Einstein-Gauss-Bonnet theory with a single maximally symmetric vacuum.
Introducing a non-minimal coupling between the scalar probe and the Ricci
scalar remarkably leads to a fully resonant spectrum for a scalar
field fulfilling reflective boundary conditions at both infinities. Applying
perturbative methods, which are particularly useful for unveiling the dynamics
at time scales of order $\varepsilon^{-2}$ (where $\varepsilon$ characterizes
the amplitude of the initial perturbation), we observe both direct and inverse energy
cascades between modes. This motivates us to explore the energy returns in the
case in which the dynamics is dominated by a single mode. We find numerical and perturbative evidence
that near exact returns do exist in this regime. We also provide some comments on
the fully backreracting case and provide a proof of the universality of the weakly non-linear dynamics around AdS, in the context of Lovelock theories with generic couplings, up to times of order $\varepsilon^{-2}$.

\end{abstract}
\maketitle

\section{Introduction}

General Relativity in dimensions higher than four can be extended, still fulfilling the requirements of second-order field
equations and diffeomorphism invariance. In general, in dimensions $D\geq5$, precise
combination of higher curvature terms can be added to the Einstein-Hilbert
action leading also to second order field equations \cite{LOVELOCK}. These
combinations are dimensional continuations of the Euler densities of the lower,
even dimensions. These combinations may appear as low energy
effective actions in string theory, as it is the case for the $\mathcal{R}^2$ term in the Heterotic and
Bosonic string theories \cite{Zwiebach:1985uq}. The simplest deformation from GR is obtained in five
dimensions, where the Einstein-Gauss-Bonnet theory has the following action
principle%
\begin{equation}
I \ [g_{\mu \nu}] = \frac{1}{16\pi G_5} \int  \left[ R- 2\Lambda + \alpha \left( R^2 - 4R_{\mu \nu} R^{\mu \nu} + R_{\mu \nu \gamma \sigma}R^{\mu \nu \gamma \sigma}  \right) \right] \sqrt{-g}d^5 x,
\end{equation} 
Note that the coupling $\alpha$ has mass dimension $-2$, and while here it
represents a free coupling.

For generic values of the couplings, the theory admits a local Lorentz
invariance which can be made manifest in the first order formulation where the
vielbein and the spin connection transform as a vector and connection of
$SO\left(  4,1\right)$, respectively. When $\alpha \Lambda = -3/4$, the
local symmetry group is enlarged to $SO\left(  4,2\right)  $ \cite{MZ},
the theory admits a unique maximally symmetric AdS solution and it has the
maximum number of propagating degrees of freedom \cite{Troncoso:1999pk}. At this
particular point, the space of solutions is also enlarged and contains, in addition to black holes \cite{Banados:1993ur} and analytic rotating solutions \cite{Anabalon:2009kq}, \cite{Anabalon:2010ns},
asymptotically locally AdS wormholes \cite{DOTTI}. The line element of the wormhole metric
reads%
\begin{equation}
ds^{2}=\ell^{2}\left[  -\cosh^{2}\rho dt^{2}+d\rho^{2}+\cosh^{2}\rho\left(
d\varphi^{2}+d\Sigma_{2}^{2}\right)  \right]  \ ,\label{wormhole}%
\end{equation}
where $-\infty<t<+\infty$, $-\infty<\rho<+\infty$, $0<\varphi\leq2\pi$ and
$d\Sigma_{2}$ stands for the line element of a compact, smooth quotient of the
pseudo-sphere with radius $3^{-1/2}$. Here $\ell^2 = 4\alpha = -3/\Lambda$. The two asymptotically locally AdS$_5$
regions $\rho\rightarrow\pm\infty$, are connected by a traversable throat
located at $\rho=0$ and the spacetime is symmetric under the reflection
$\rho\rightarrow-\rho$. This spacetime being devoid of singularities and
horizons, represents a soliton in the non-linear
Einstein-Gauss-Bonnet theory \footnote{It is interesting to notice that a similar situation occurs for Einsteinian gravities \cite{Bueno:2018uoy}, where it was shown that a wormhole exists for a particular value of the coupling constants. See also \cite{ACZ} for Lorentzian wormholes in GR supported by a Skyrme field.}.

The propagation of a scalar probe on the background geometry (\ref{wormhole}) was originally explored in \cite{COT}, where the focus was on the computation, in a closed form, of the normal frequencies fulfilling different possible boundary conditions. In \cite{NOSOTROS}, this problem was partially revisited and it was shown that for a particular value of a non-minimal coupling with the scalar curvature, the propagation of the scalar is controlled by an effective Schr\"{o}dinger problem in a
Rosen-Morse potential, for which the energies are proportional to the square
of the frequencies of the scalar probe. Since the eigenvalues of the
Schr\"{o}dinger operator in Rosen-Morse potentials are quadratic in the mode
number $n$ \cite{cooperreview}, the spectrum of the purely radial scalar probe turn
out to be equispaced or fully resonant. Equispaced spectra play an important role in turbulent energy transfer leading to the non-perturbative AdS instability \cite{Turbulent1}-\cite{Turbulent6}. This leads to a rich phenomenology that also appears in non-linear models of different physical nature as in self-gravitating scalars on a spherical cavity in 3+1
\cite{MALIBORSKI}, on systems describing Bose-Einstein condensates
\cite{BIASI-2018}, and vortex precession
\cite{Biasi:2017pdp}, as well as in the conformal dynamics on the Einstein
Universe \cite{Bizon:2016uyg}. A program to classify all the spacetimes in
which a minimally coupled scalar probe may lead to an exactly solvable  effective Schroedinger problem, suggestively
dubbed \textquotedblleft Klein-Gordonization", was initiated in \cite{KLEIN}.

\bigskip

In this paper we go beyond the linear level and perturbativelly explore different aspects of the non-linear
dynamics of a fully resonant, self-interacting scalar
probe on the wormhole spacetime (\ref{wormhole}).

In Section II, we revisit with detail the linear propagation of a scalar probe with a precise non-minimal coupling with
the Ricci scalar. The solutions to this linear problem are dubbed ``wormhole oscillons" and we show that the spectrum of these oscillons is linear in the mode number. After introducing a self-interaction, in Section III we
construct the system of equations that control the dynamics of the infinite
oscillators in the Two-Time-Framework (TTF) \cite{holotherm} or the time averaged
system \cite{Craps:2014vaa}-\cite{Craps:2014jwa}. This approach has been particularly useful in the
context of the non-perturbative instability of AdS, since it captures the
dynamics at times of order $\varepsilon^{-2}$, where $\varepsilon$
characterizes the energy content of the initial perturbation. By truncating
the system of oscillators we study the energy transfer between modes, and
show that there are direct and inverse energy cascades. In particular, when
the dynamics is dominated by a single mode, we find evidence of near exact
energy returns, which is confirmed in Section IV analytically using perturbation theory. Section V is
devoted to conclusions and further comments on the universality of the weakly non-linear dynamics on AdS for Lovelock theories with generic couplings. It is well-known that higher curvature gravity theories may have more than one maximally symmetric vacuum. We show that for an arbitrary Lovelock theory, provided the couplings are generic, the form of the equation for the infinite oscillators that control the dynamics in the TTF is universal.

\section{The linear scalar probe}

Let us consider the equation for a scalar probe on the wormhole geometry
(\ref{wormhole}), with a fixed non-minimal coupling\footnote{Note that this is not the conformal coupling since in general $\xi_{\mathrm{conf}}=\frac{D-2}{4(D-1)}$.}%
\begin{equation}
\left(  \square-m^{2}-\frac{3}{8}R\right)  \phi_1\left(  x^{\mu}\right)  =0\ .
\label{waveequation}
\end{equation}
We will see below that even though the Ricci scalar is a non-trivial function of the radial coordinate
\begin{equation}
    R = -\frac{20}{\ell^2} + \frac{6}{\ell^2 \cosh^2(\rho)} ,
\end{equation}
the equation for the scalar probe can be solved analytically. Hereafter, for simplicity, we fix $\ell=1$.

Introducing a mode separation and considering only a radial spatial dependence 
\begin{eqnarray}
    \phi_1(t,\rho) &=& e^{-i\omega t} R(\rho) \ ,
\end{eqnarray}
and the radial coordinate $\rho^{*} = 2 \arctan(e^{\rho})$ which maps $\rho \in \ ]\-\infty,\infty[$ to $\rho^{*} \in \ ]0,\pi[$, we obtain
    \begin{equation}
    - \frac{d^2 S(\rho^{*})}{d \rho^{*2}} + U (\rho^{*}) S (\rho^{*}) = \omega^2 S(\rho^{*}) \label{schr}
    \end{equation}
    where $S(\rho)=\cosh(\rho)^{3/2}R(\rho)$ and the effective potential reads
    \begin{equation}
        U(\rho^{*}):=\frac{1}{4} \ \frac{4 m^2 - 15}{\sin^2(\rho^{*})}.
    \end{equation}
In terms of the coordinate $\rho^{*}$, the metric is manifestly conformal to the product of $\mathbb{R}_t\times\mathbb{R}\times S^1\times \Sigma_{2}$, and reads
\begin{equation}
ds^2=\frac{1}{\sin^2(\rho^{*})}[-dt^2+d\rho^{*2}+d\varphi^2+d\Sigma_2^2]\ ,
\end{equation}
and the wormhole boundaries are located at the divergences of the conformal factor.

The equation \eqref{schr} is that of a quantum particle moving in a Rosen-Morse potential. For the following analysis it is convenient to introduce the coordinate $z$ such that 
\begin{eqnarray} 
    \tanh(\rho) &=& 1-2z \ ,
\end{eqnarray}
which maps $\rho \in \ ]\-\infty,\infty[$ to $z \in \  ]1,0[$. The wave equation \eqref{waveequation} leads to
\begin{eqnarray}\label{eqphi1wh}
 \ddot{\phi_1} + L \phi_1 = 0
\end{eqnarray}
where the operator $L$, is defined as
\begin{eqnarray}
L &:=& -z^2(1-z)^2 \frac{d}{dz} \left(\frac{1}{z(1-z)} \frac{d}{dz} \right) + \frac{m^2-15/2}{4 z (1-z)} + \frac{9}{4}. \label{Lwhfinal}
\end{eqnarray}
This operator admits the following asymptotic behaviors
    \begin{align}
    \begin{split}
    R(z) &\overset{z \rightarrow 0}{\sim} D_1 \  z^{\Delta_+} + D_2 \  z^{\Delta_-}, \\
    R(z) &\overset{z \rightarrow 1}{\sim} \tilde{D}_1 (1-z)^{\Delta_+} + \tilde{D}_2 (1-z)^{\Delta_-},
    \end{split}\label{dondeestanlosdeltas}
    \end{align}
where $\Delta_{\pm}= 1 \pm \frac{1}{2}\sqrt{m^2 - \frac{7}{2}}$. Note that $\Delta_+>0$, while $\Delta_-\leq 0$ for $m^2\geq 15/2$. Assuming $m^2\geq 15/2$ (which in global AdS would correspond to $m^2\geq 0$), we impose reflective boundary conditions at both boundaries $z=1$ and $z=0$, setting $D_2=\tilde{D}_2=0$, and consequently the operator $L$ is essentially self-adjoint on $L^2([1,0],-\frac{1}{z^2(1-z)^2})$. This differential eigenvalue problem therefore leads to the following normal frequencies and normal modes
\begin{eqnarray}
\omega_j^2 &=& \left( j + \frac{1}{2} + \sqrt{
m^2 - \frac{7}{2}}\right)^2 \ , \label{autovaloreswh} \\
\nonumber e_j(z) &=& C_j \ z^{1+\frac{1}{2}\sqrt{m^2-\frac{7}{2}}} (1-z)^{1-\frac{1}{2}\sqrt{m^2-\frac{7}{2}}}  \\
&& \times \left._2 F_1 \right. \left(-j -\sqrt{m^2-\frac{7}{2}},1+j+\sqrt{m^2-\frac{7}{2}};1+\sqrt{m^2-\frac{7}{2}};z \right)\ .\label{autofuncioneswh}
\end{eqnarray}
The latter can also be written in terms of Jacobi Polynomials $P_{n}^{(a,b)}(1-2z)$, such that
\begin{eqnarray}
e_j(z)&=&D_{j}z^{1+\frac{1}{2}\sqrt{m^{2}-\frac{7}{2}}%
}\left(  1-z\right)  ^{1-\frac{1}{2}\sqrt{m^{2}-\frac{7}{2}}} P_{j+\sqrt{m^{2}-\frac{7}{2}}}^{\left(  \sqrt{m^{2}-\frac{7}{2}%
},-\sqrt{m^{2}-\frac{7}{2}}\right)  }\left(  1-2z\right)\label{autofuncioneswhJac}
\end{eqnarray}
where $C_j$ and $D_j$  are normalization constants that depend on the mass of the scalar probe, and are proportional, the proportionality factor being a quotient of Gamma functions. Hereafter we refer to \eqref{autofuncioneswh} as ``wormhole oscillons".  The general solution to the linear problem is given by an arbitrary superposition of the modes \eqref{autofuncioneswh}, leading to
\begin{equation}
    \phi_1(t,z) = \sum_{j=0}^{\infty} a_j \cos(\omega_j t + \beta_j) e_j(z) .
\end{equation}
For simplicity, and mimicking the massless case in AdS, we set $m^2=15/2$. In this case the normalization constants in \eqref{autofuncioneswh} fulfil
\begin{equation}
2C_j=\left[(j+1)(j+2)(j+3)(j+4)(5+2j)\right]^{1/2}
\end{equation}
leading to $(e_i(z),e_j(z))=\delta_{ij}$. The frequencies which were already equispaced in \eqref{autovaloreswh}, further reduce to
\begin{equation}
\omega_j=\frac{5}{2}+j
\end{equation}
with $j=0,1,2,...$ .

Note that we have been able to find a fully resonant, equispaced spectrum for a scalar probe propagating on a spacetime with non-trivial topology. Below, we introduce a self-interaction on the scalar to characterize the energy transfer between modes. As mentioned above, it has been shown that such problem captures some features of the backreacting, massless scalar in AdS (see e.g. \cite{selfint1}, \cite{Basu:2015efa}).

\section{Self-interacting scalar probe}
Now we will introduce a non-linearity in the scalar probe we discussed in the previous section. In particular we will focus on
\begin{equation}\label{eqphiselfwh}
    \Box \phi - \left(m^2 +\frac{3}{8} R\right)\phi - \frac{\lambda}{3!}\phi^3 = 0 \ ,
\end{equation} 
setting $m^2=15/2$, since this value leads to $\Delta_{-}=0$ in \eqref{dondeestanlosdeltas} and mimics the massless case in AdS. Here $\lambda$ is a constant with mass dimension $-1$. Following the TTF we introduce the slow time $\tau=\epsilon^2 t$ and the perturbative ansatz
\begin{eqnarray}
\phi(t,\tau,z) &=& \sum_{j=0}^{\infty} \epsilon^{2j+1} \phi_{2j+1} (t,\tau,z)\ .  \label{expphiwh} 
\end{eqnarray}
Note that a direct perturbative approach leads to resonant terms, some of which could be perturbatively absorbed by a Poincare-Lindstedt shift. The TTF helps dealing with this feature, and even more its validity is ensured at least up to times of order $\epsilon^{-2}$. Naturally, at first order in $\epsilon$ one re-obtains the linear problem
\begin{equation}\label{eqphi1TTFwh}
\partial^2_t \phi_1 + L \phi_1 = 0,
\end{equation}
where the operator $L$ is given in \eqref{Lwhfinal}, leading to
\begin{equation}
\phi_1(t,\tau,x) = \sum_{l=0}^{\infty} \left( A_l (\tau) e^{-i \omega_l t} + \bar{A}_l (\tau) e^{i \omega_l t} \right) e_l (x), 
\end{equation}
where $A_l(\tau)$ are arbitrary functions of the slow time $\tau$. Here $\bar{A}_l(\tau)$ stands for the complex conjugate of $A_l(\tau)$. At the next perturbative order in $\epsilon$ one obtains
\begin{equation}\label{eqphi3ttfwh}
\partial_t^2 \phi_3 + L \phi_3 + 2 \partial_t \partial_{\tau} \phi_1 = S(t,\tau,x),
\end{equation}
with the source given by the lower order term $S=-\frac{1}{24}\frac{\phi_1^3}{z(1-z)}$. Here one proposes a solution for $\phi_3$ of the form:
\begin{equation}
\phi_3(t,\tau,x) = \sum_n^{\infty} \left( B_n(t,\tau) + \bar{B}_n(t,\tau) \right) e_n(x).
\end{equation}
Projecting equation \eqref{eqphi3ttfwh} on the basis of wormhole oscillons one obtains
\begin{equation}
\partial_t^2 \left( B_j + \bar{B}_j \right) + \omega^2_j \left(B_j + \bar{B}_j \right) - 2 i \omega_j(\partial_{\tau} A_j e^{-i \omega_j t} - \partial_{\tau} \bar{A}_j e^{i \omega_j t}) = (e_j,S)\ ,\label{conlosBs}
\end{equation}
with
\begin{eqnarray}
\nonumber (e_j,S)&=&  \sum_{n,l,m} \frac{\mathcal{S}_{jnlm}}{3} \left(  A_l  A_m A_n e^{-i (\omega_l + \omega_m + \omega_n) t} + 3  A_l  A_m \bar{A}_n e^{-i (\omega_l + \omega_m - \omega_n) t} \right. \\
\nonumber && \left. \ \ \ \ \ \ \ \ \ \ \  + 3  A_l   \bar{A}_m  \bar{A}_n  e^{-i (\omega_l - \omega_m - \omega_n) t} +   \bar{A}_l  \bar{A}_m \bar{A}_n e^{-i (-\omega_l - \omega_m - \omega_n) t} \right), \\
&& 
\end{eqnarray}
where the interaction integrals $\mathcal{S}_{jnlm}\in \mathbb{R}$ are defined as
\begin{equation}
\mathcal{S}_{jnlm} = \frac{1}{8} \int_{1}^{0} \frac{e_j(z) e_n(z) e_l(z) e_m(z)}{z^3 (1-z)^3} dz \ .\label{overlapintegrals}
\end{equation}
The TTF equations are obtained by imposing that the functions $A(\tau)$ that appear in the l.h.s. of equation \eqref{conlosBs}, exactly cancel the resonant terms coming from the r.h.s. of the same equation. This leads to
\begin{eqnarray}
\nonumber -2 i \omega_j \partial_{\tau} A_j &=& \sum_{n,l,m} \frac{\mathcal{S}_{jnlm}}{3} \left[A_l  A_m  A_n \delta_{\omega_j, \omega_l + \omega_m + \omega_n} + 3  A_l  A_m \bar{A}_n \delta_{\omega_j,\omega_l + \omega_m - \omega_n} \right. \\
&& \ \ \ \ \ \ \ \ \ \ \left.  + 3  A_l  \bar{A}_m \bar{A}_n \delta_{\omega_j, \omega_l - \omega_m - \omega_n} +   \bar{A}_l  \bar{A}_m  \bar{A}_n \delta_{\omega_j,-\omega_l - \omega_m - \omega_n} \right]\ .
\end{eqnarray}
We observe that the integrals $\mathcal{S}_{jnlm}$, are non-vanishing only in the channel $\omega_j+\omega_n=\omega_l+\omega_m$ (or equivalently $j+n=l+m$), leading to the TTF equations
\begin{equation}
 -2 i \omega_j \partial_{\tau} A_j  = \sum_{j+n=m+l} \mathcal{S}_{jnlm} A_l  A_m  \bar{A}_n \ . \label{TTFeqs}
\end{equation}
The vanishing of the channels $\omega_j=\omega_l+\omega_m+\omega_n$ and $\omega_j=\omega_l-\omega_m-\omega_n$ is a special property of our equispaced spectrum, as in AdS. The vanishing of the overlap integrals $\mathcal{S}$ in these cases can be proved using the expressions for the wormhole oscillons in terms of Jacobi polynomials in \eqref{autofuncioneswhJac}, as for a scalar probe in AdS \cite{vanres1}-\cite{vanres2}. Observing the symmetries in the indices of the overlap integrals $\mathcal{S}_{jnlm}$, and the fact they vanish unless $j+n=m+l$, one can deduce that the total energy $E=\sum_j\omega_j^2\bar A_j(\tau)A_j(\tau)$ and the ``particle number" $N=\sum_j\omega_j\bar A_j(\tau)A_j(\tau)$ are conserved \cite{Craps:2014jwa}, \cite{selfint1}, \cite{CASCADAS}. The conservation of these quantities is particularly useful for monitoring the stability of the numerical integration of the truncated version of the system \eqref{TTFeqs}.

In what follows we will solve the system of oscillators by truncating the sum up to order $j=j_{\mathrm{max}}$, for different initial data. We monitor the convergence by increasing $j_{\mathrm{max}}$ and study the energy transfer between modes induced by the non-linearities.

\begin{figure}[h!]
\includegraphics[scale=0.67]{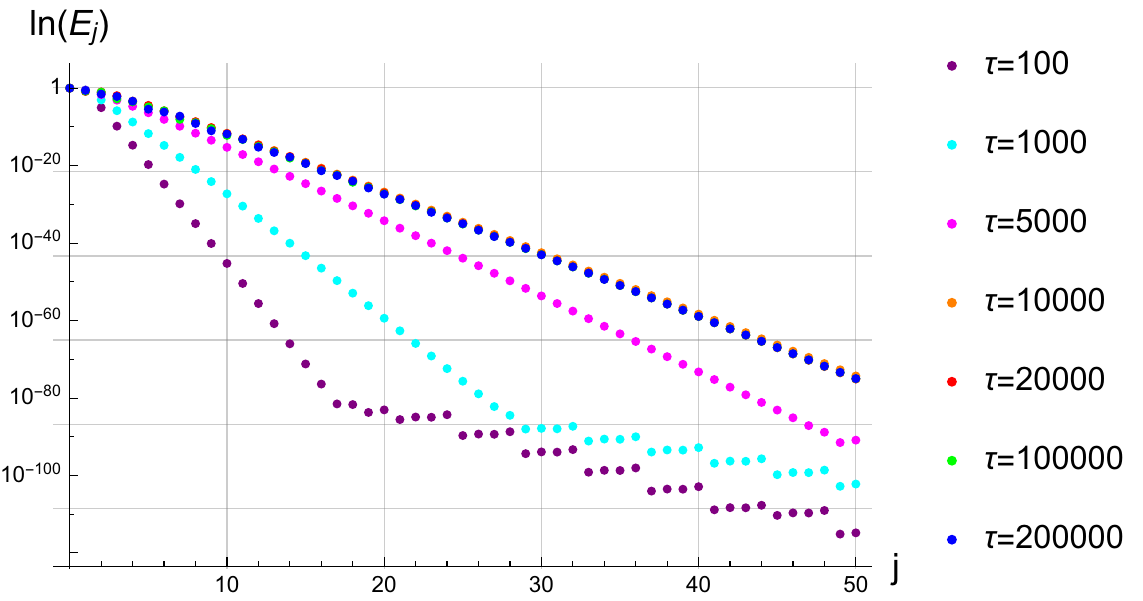}\qquad
\includegraphics[scale=0.67]{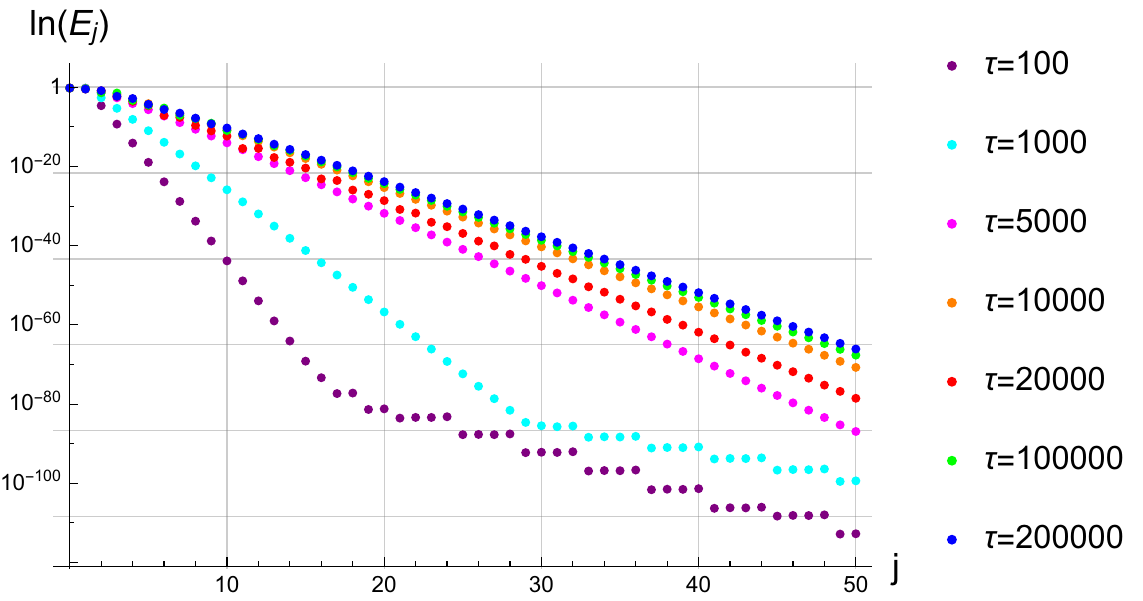}\caption{Evolution of the energy per mode as a function of the mode number $j$, for $(E_0(0),E_1(0))=(3/4,1/4)$ (left-panel) and $(E_0(0),E_1(0))=(1/2,1/2)$ (right-panel). For late times the energy per mode is exponentially suppressed for large $j$, i.e. $E_j\sim e^{-j}$.}%
\label{Eversusj}%
\end{figure}

\begin{figure}[h!]
\includegraphics[scale=0.7]{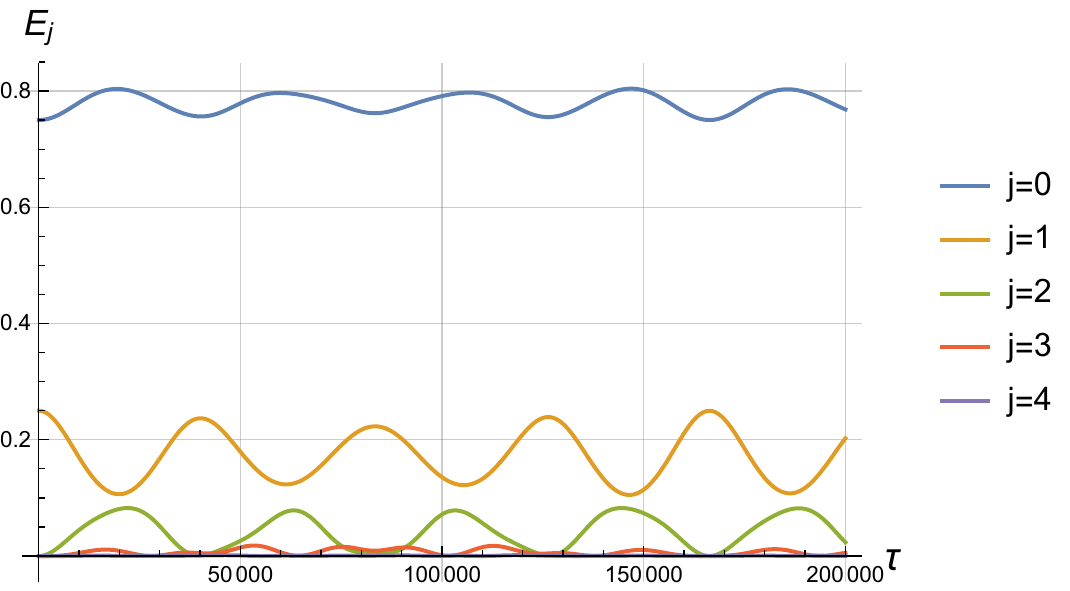}\qquad
\includegraphics[scale=0.7]{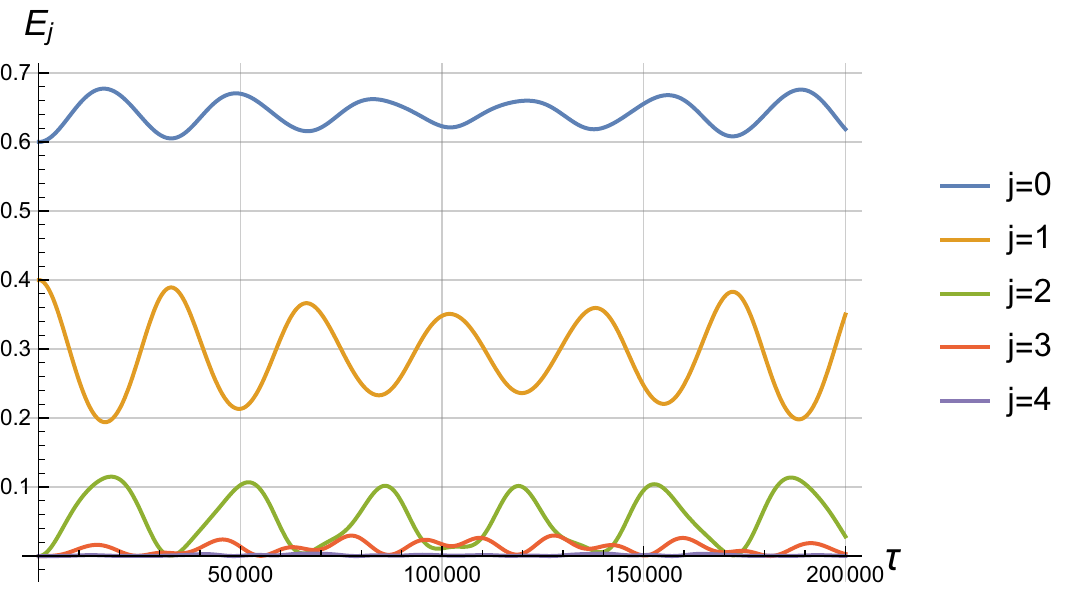}\qquad
\includegraphics[scale=0.7]{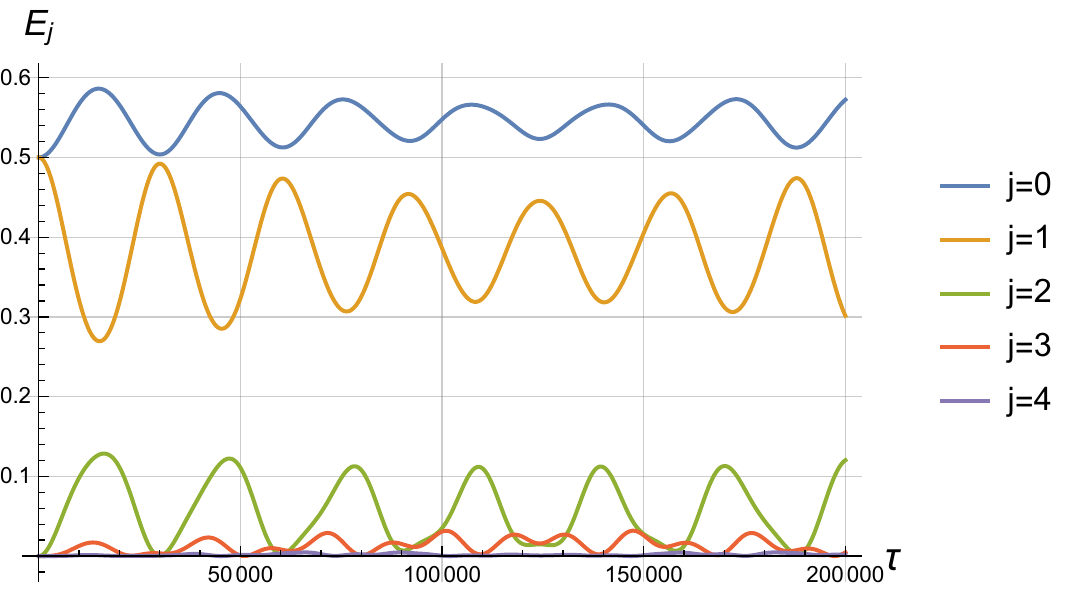}\qquad
\includegraphics[scale=0.7]{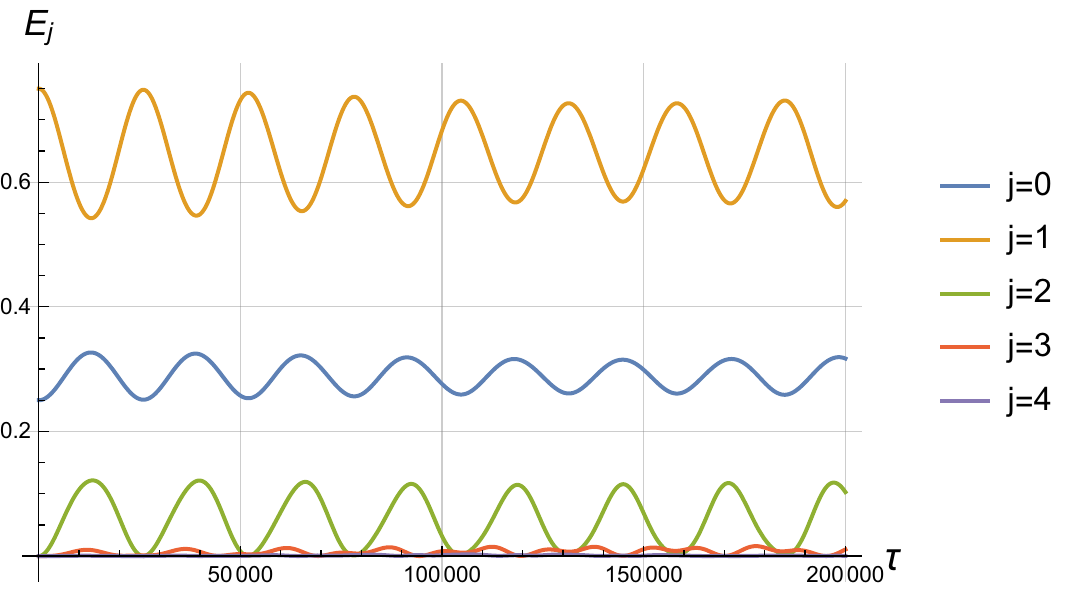}\caption{The plots present the evolution of the energy per mode, for different initial excitations in the fundamental and first excited mode. The upper left panel corresponds to $(E_0(0),E_1(0))=(3/4,1/4)$, for the upper right panel we have used $(E_0(0),E_1(0))=(3/5,2/5)$, the lower left panel corresponds to the two-mode equal energy initial date $(E_0(0),E_1(0))=(1/2,1/2)$ and finally, for the lower right panel $(E_0(0),E_1(0))=(1/4,3/4)$.}%
\label{Eversustau}%
\end{figure}

In Figure 1 we plot the evolution of the spectrum, showing energy transfer induced by the non-linearities, for different initial conditions. We have evolved the truncated TTF system with $j_{\mathrm{max}}=50$. The spectra stabilize after some time, showing an exponential suppression of the energy as a function of the mode number. As it occurs for non-backreacting probes in AdS, these spectra suggest the absence of a turbulent phenomenology\footnote{See e.g. \cite{deOliveira:2012dt} for a turbulent characterization of the power spectrum in $D=4,5$ in GR}.

Figure 2 shows the actual time evolution of the energy per mode, for different initial conditions with $j_{\mathrm{max}}=50$. Even though the energy is initially distributed only in the fundamental and first excited modes, the non-linearities transfer energy to the higher harmonics. The plots suggest energy returns after a finite time. In the next section we provide perturbative evidence of near exact energy returns for situations as that depicted in the upper left panel, in which the dynamics is clearly dominated by the fundamental mode.

It is also illustrative to consider initial data with three and four modes turned on. Figure 3 shows the time evolution of the energy content as well as the evolution of the stabilized spectra. Note that also in this case, the energy in modes with large $j$ are exponentially suppressed.

\begin{figure}[h!]
\includegraphics[scale=0.65]{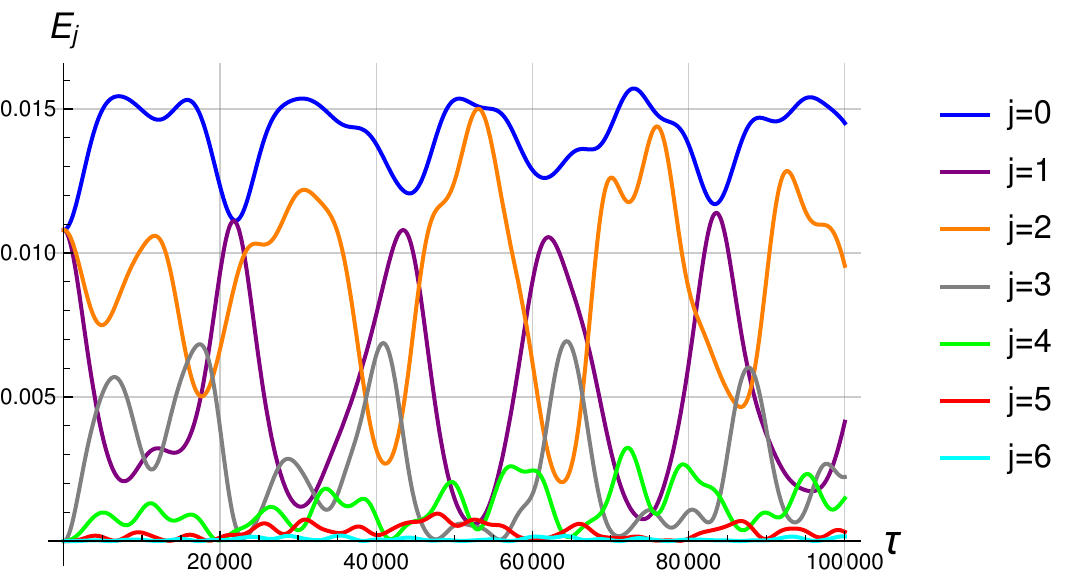}\qquad
\includegraphics[scale=0.65]{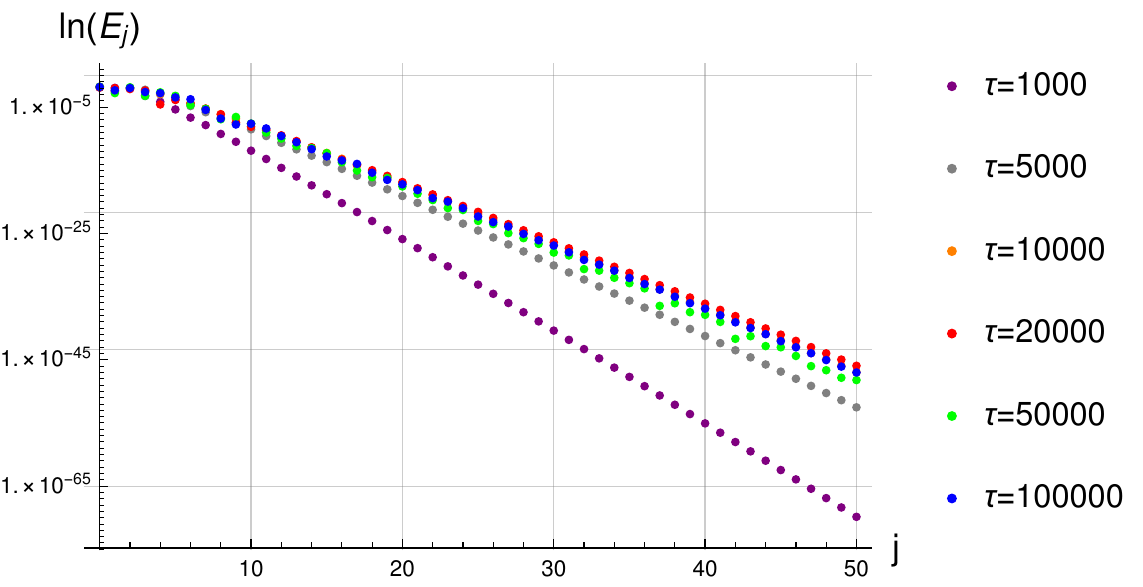}\qquad
\includegraphics[scale=0.65]{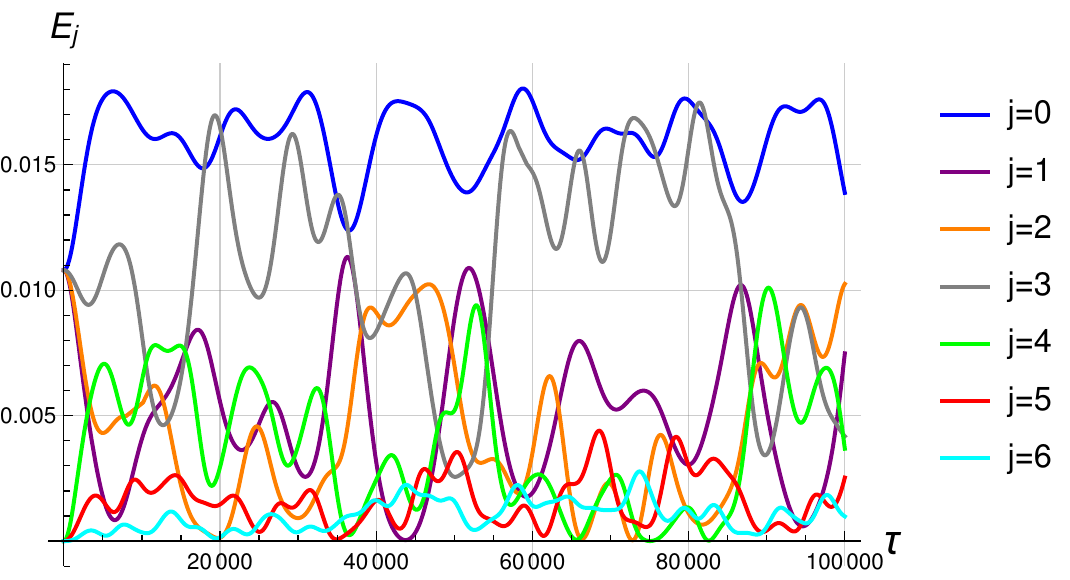}\qquad
\includegraphics[scale=0.65]{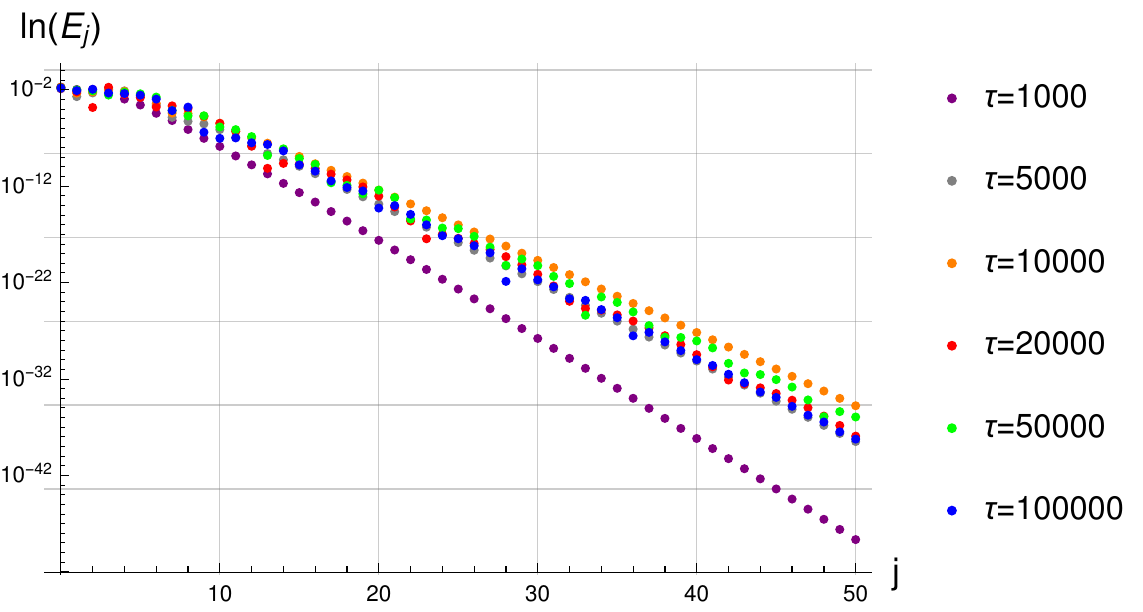}\caption{Time evolution of the energy and spectra for three (upper panel) and four (lower panel) modes with equal energy as initial conditions.}%
\label{threefour}%
\end{figure}

\section{Near exact energy returns}
Closely following reference \cite{RETURN}, a perturbative argument can be
given to have an analytic understanding of the (near) exact energy returns suggested by
Figure 2. We will focus on the situation such that the dynamics is dominated by the fundamental mode, as well as in the case where the first excited mode dominates.
\subsection{Fundamental mode dominating the dynamics}
In particular in the case in which the dynamics is dominated by the
fundamental mode, we can introduce the scaled oscillators $q_j$ such that%
\begin{equation}
A_{j}\left(  \tau\right)  =\frac{q_{j}\left(  \tau\right)  }{\sqrt{\omega_{j}%
}}\delta^{j}\ ,
\end{equation}
with $\delta$ a small, perturbative parameter. The time-averaged system now reads%
\begin{equation}
i\partial_{\tau}q_{j}=\sum_{m=0}^{+\infty}\sum_{k=0}^{j+m}\delta
^{2m}C_{j,m,k,j+m-k}q_{k}q_{n+m-k}\bar{q}_{m}\ ,
\end{equation}
where%
\begin{equation}
C_{j,n,l,m}=-\frac{1}{2}\frac{\mathcal{S}_{jnlm}}{\sqrt{\omega_{j}\omega
_{n}\omega_{l}\omega_{m}}}\ ,\label{scaledSs}
\end{equation}
and the overlap integrals have been defined in \eqref{overlapintegrals}. At leading order in $\delta$ we obtain the non-linear system%
\begin{equation}
i\partial_{\tau}q_{j}=\bar{q}_{0}\sum_{k=0}^{j}C_{j,0,k,j-k}q_{k}q_{n-k}\ .
\end{equation}
For $j=0$, the system leads to a decoupled, non-linear equation for $q_{0}$,
which is solved in a closed form, giving a constant modulus and a time
dependent phase for $q_{0}\in \mathbb{C}$. The global symmetries of the system can
be used to set the absolute value of $q_{0}$ to $1$. Then, for $j\geq1$, one
obtains a set of linear equations that can be solved in a recursive manner,
where the $q_{k<j}$'s appear as sources. The homogeneous equations depend only
on the coefficients of the form $C_{j0j0}$ (Figure 4 depicts these integrals
up to $j\sim500$). A further use of the symmetries of the system allows to set%
\begin{equation}
q_{0}\left(  \tau\right)  =e^{-iC_{0000}\tau}\text{ and }q_{1}\left(
\tau\right)  =e^{-2iC_{1010}\tau}\ .
\end{equation}

\begin{figure}[h!]
\includegraphics[scale=0.6]{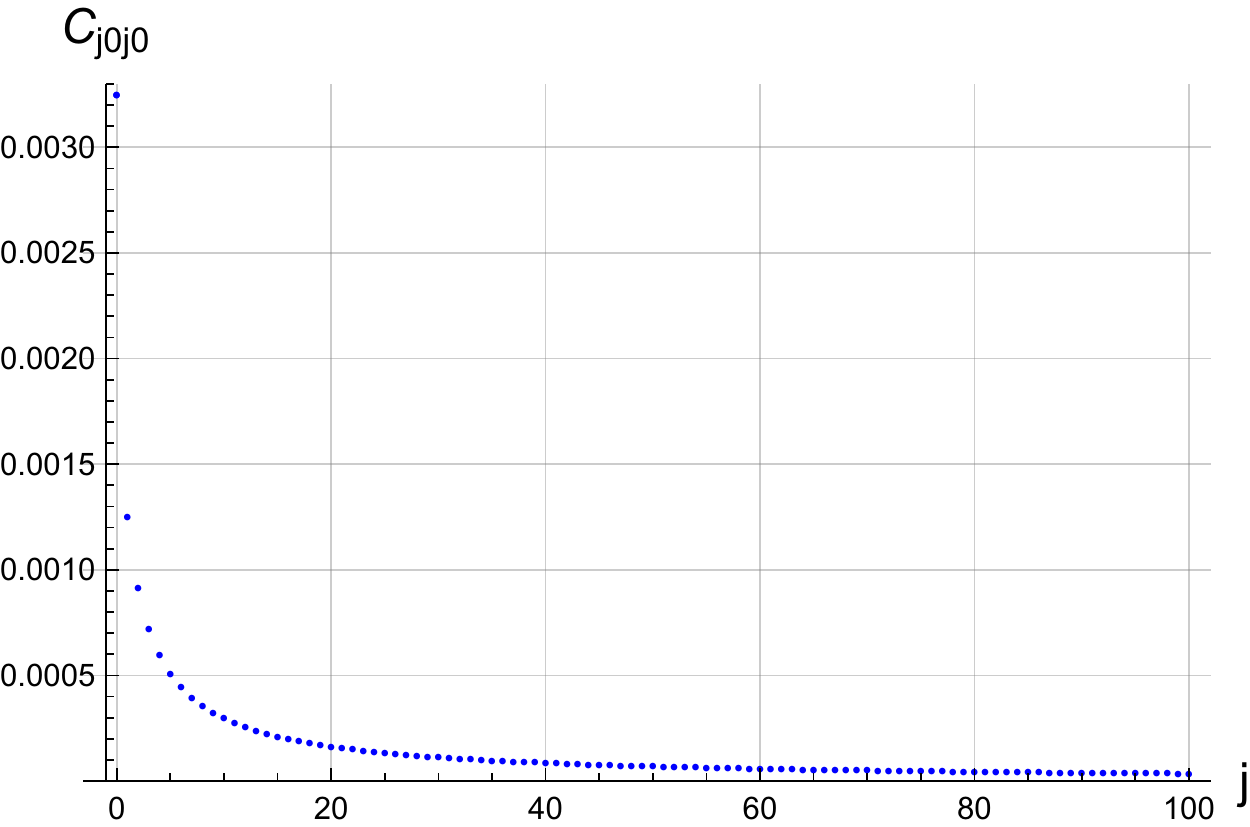}\qquad
\includegraphics[scale=0.6]{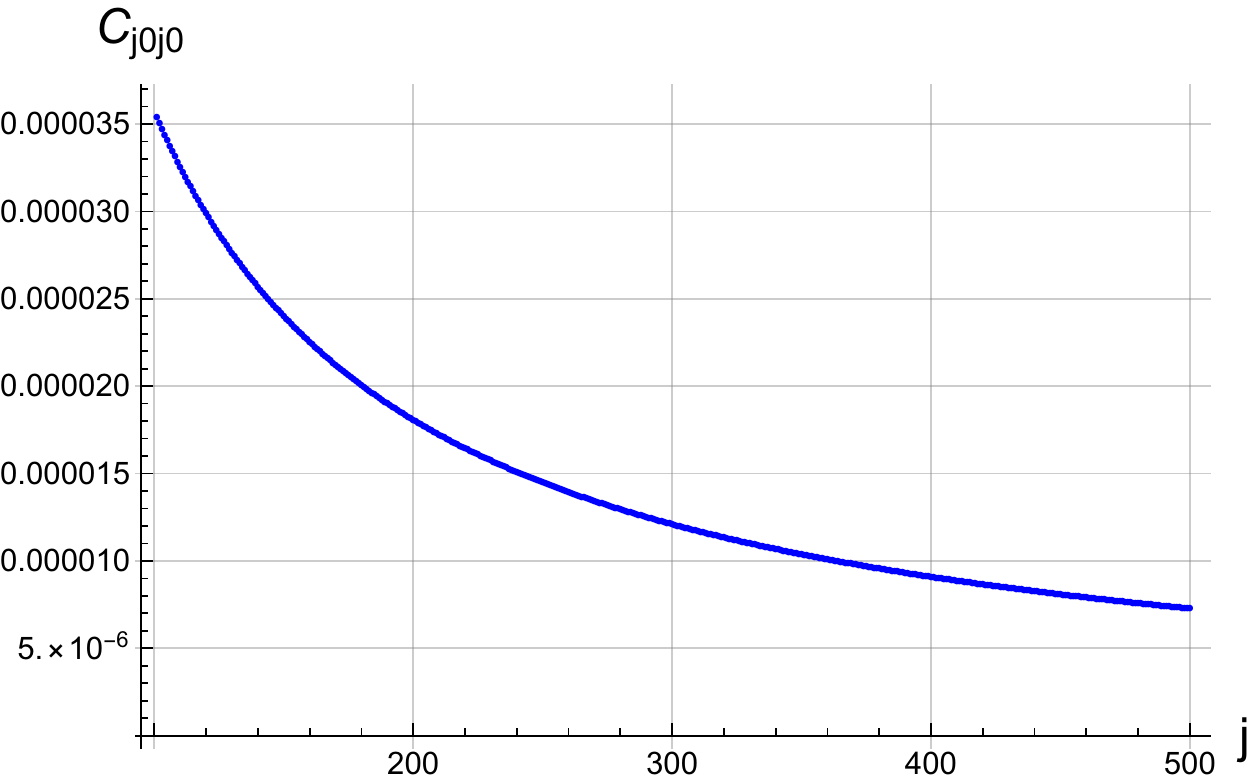}\caption{The scaled overlap integral $C_{j0j0}$ that determine the dynamics when the fundamental mode dominates.}%
\label{Eversustau}%
\end{figure}

With this in mind one can compute the time periods of the energies in the
higher modes by computing the periods $T_{j}$ of $E_{j}\sim q_{j}\bar{q}_{j}$.
Using our overlap integrals we obtain $T_{2}=24024\pi$
and $
T_{3}=17T_{2},\text{ }T_{4}=19T_{3},\text{ }T_{5}=T_{4}\text{, }T_{6}%
=23T_{5}$.
Note that the ratios of the frequencies are relatively simple fractions (a simple, pictorial method to see the exact and near exact returns is outlined in Figure 5). The commensurability of the periods of the energy, ensure exact energy returns at finite time within this perturbative approach. Nevertheless it must be noted that the periods $T_j$ are an increasing function of the mode number $j$, and therefore as more modes are included in the analysis one should have to wait longer for observing the recurrence. Note that higher modes are suppressed as $\delta^j$.

\begin{figure}[h!]
\includegraphics[scale=0.55]{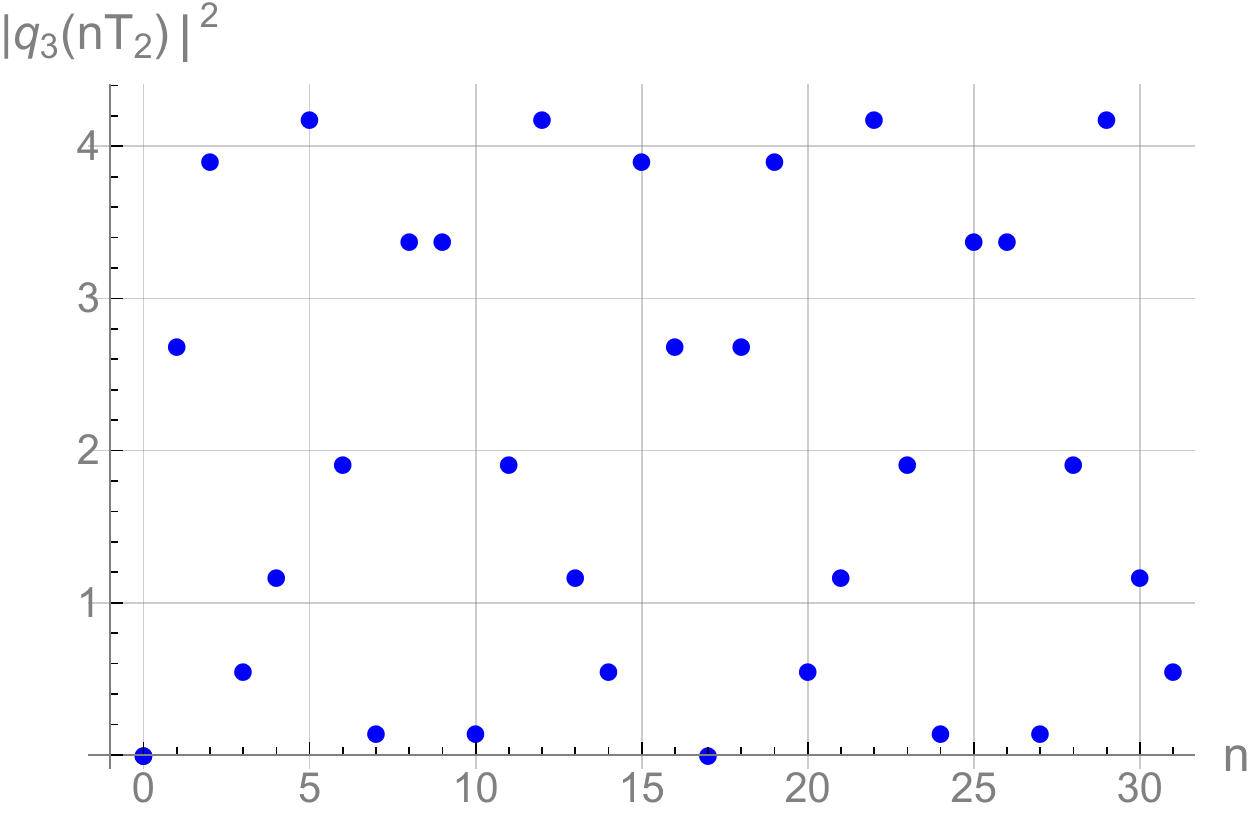}\qquad
\includegraphics[scale=0.55]{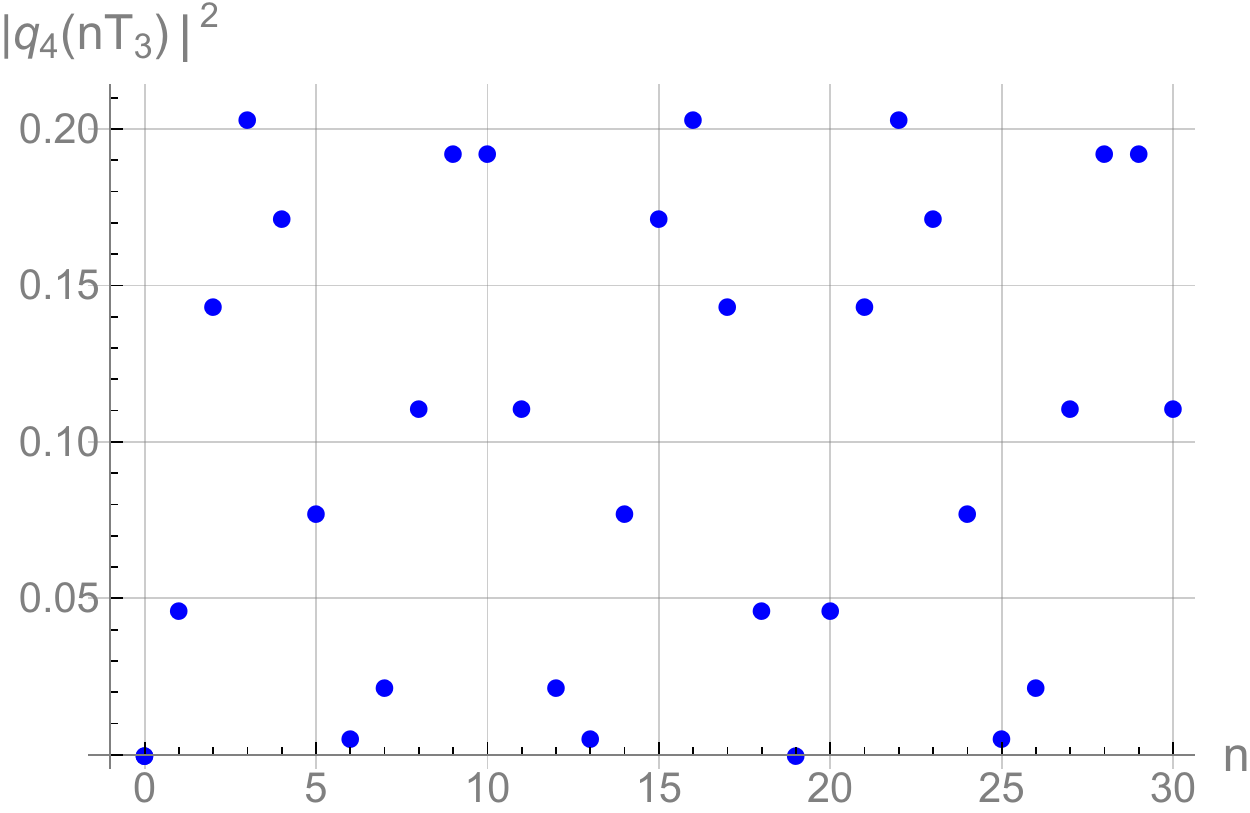}\qquad
\includegraphics[scale=0.55]{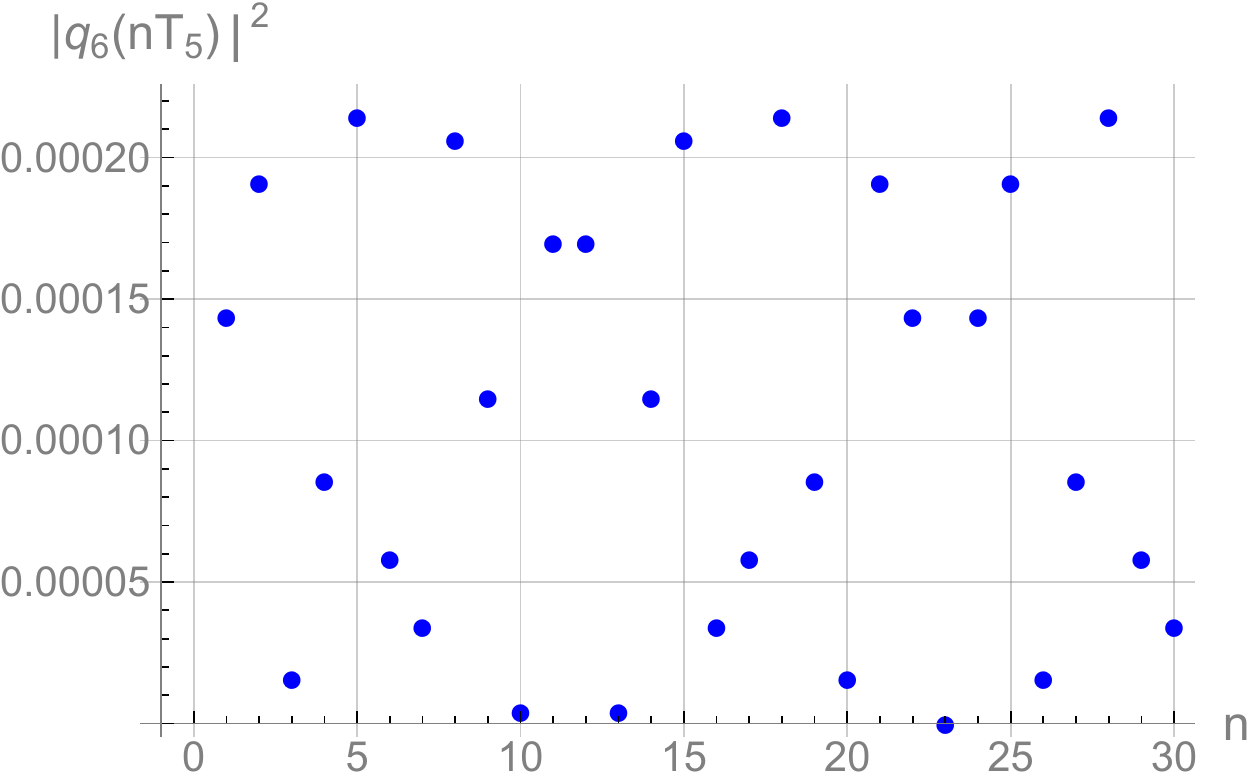}\caption{The figures depict the exact and near exact energy returns for the modes with $j=3,4$ and $6$. Since in the initial condition these modes do not have energy, the exact returns are obtained once the dots intersect the horizontal line.}%
\label{Eversustau}%
\end{figure}

\subsection{First excited mode dominating the dynamics}

The lower-right panel of Figure 2 depicts a case in which the first excited mode is dominating the dynamics of the energy content in the system. We can analytically explore such case in a perturbative manner by introducing the ansatz
\begin{equation}
A_{0}=\frac{q_{0}\left(  \tau\right)  }{\sqrt{\omega_{0}%
}}\delta \ \ \text{and} \  
A_{j\geq 1}\left(  \tau\right)  =\frac{q_{j}\left(  \tau\right)  }{\sqrt{\omega_{j}%
}}\delta^{j-1}\ .
\end{equation}
Retaining the leading contributions as $\delta$ goes to zero, one obtains
\begin{equation}
i\dot{q}_0=C_{0211}\bar{q}_2q_1^2+2C_{0101}|q_1|^2q_0\ ,
\end{equation}
and
\begin{equation}
i\dot{q}_n=\bar{q}_1 \sum_{k=1}^{n} C_{n,1,k,n+1-k}q_k q_{n+1-k}+\bar{q}_0\sum_{k=1}^{n-1}C_{n,0,k,n-k}q_k q_{n-k} .
\end{equation}
where the couplings $C$'s have been defined in \eqref{scaledSs}. As before using the global symmetries, the time dependence of the leading oscillator appear in its phase, while the time dependence of the first two subleading oscillators leads to equal periods of their energy content given by $T_0=T_2=14586\sqrt{14/635}\pi$. For the sub-subleading modes, after some manipulation (see Section 4.2 of \cite{RETURN}), one obtains that $q_3(\tau)$ is a superposition of non-commesurable oscillations (see Figure 6).

\begin{figure}[h!]
\includegraphics[scale=0.7]{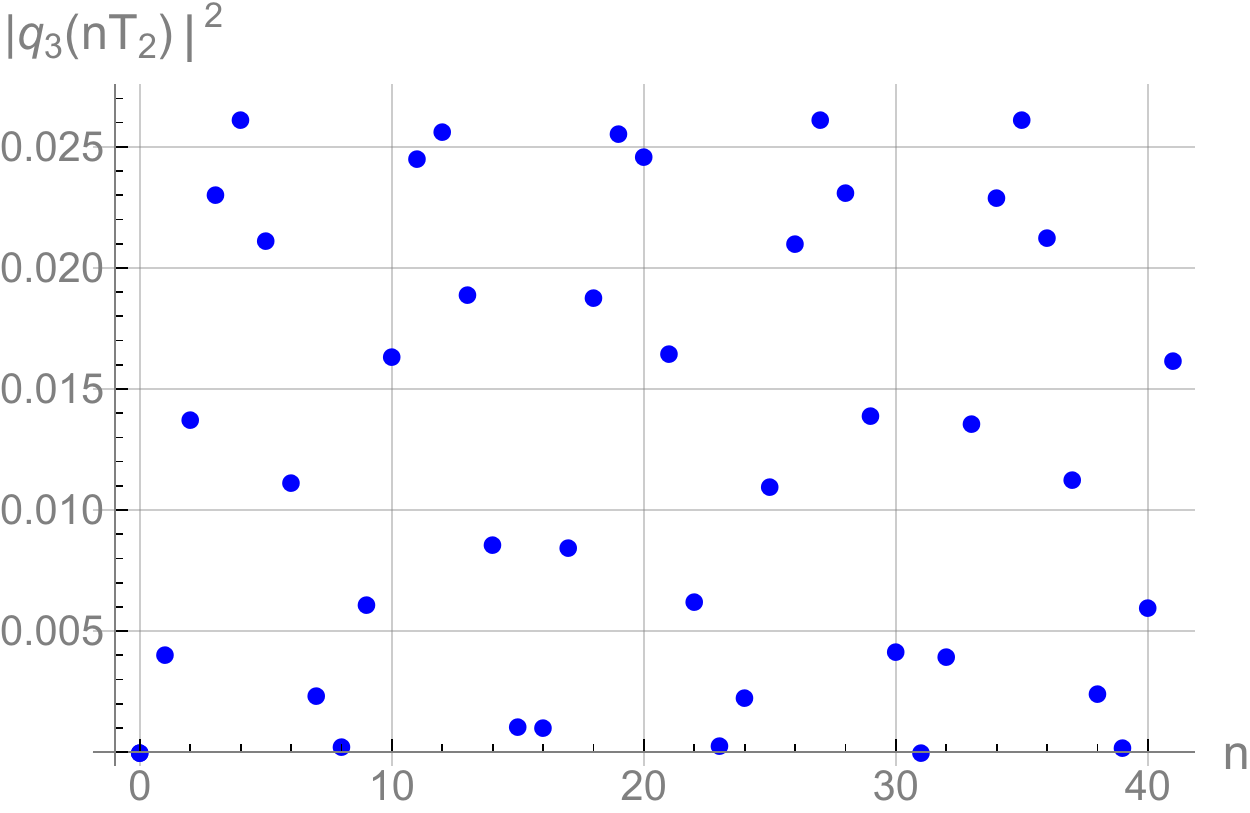}\caption{This plot exhibits the near exact returns of the energy in the third mode. Even though some point seem to lie on the horizontal axis, the actually do not touch it since for example for $n=31$, $|q_3(nT_2)|$ is of the order $10^{-7}$.}%
\label{q3domina1}%
\end{figure}

It is important to notice that in both of the cases developed in Sections A and B, the returns are only near exact due to the energy transfer to modes with higher values of $j$. Nevertheless, if one truncates the system to include only low modes, sections A and B differ. In the former, since the frequencies are commensurable, there will be near-exact as well as exact returns (of the truncated version), while in the latter, the non-commensurability of the modes will only allow for near exact returns with different levels of precision, even in this truncated version.

\section{Further comments}
In this work we have shown the existence of a spacetime with non-trivial topology on which the linear dynamics of a scalar probe turn out to be fully resonant, leading to a rich phenomenology when non-linearities are included. The five-dimensional wormhole geometry explored in this work, can be generalized to arbitrary odd dimensions $D=2n+1$ (with $n\geq 2$)
\begin{equation}
ds^{2}=\ell^{2}\left[  -\cosh^{2}\rho dt^{2}+d\rho^{2}+\cosh^{2}\rho
d\Sigma_{D-2}^{2}  \right]  \ ,\label{wormholegend}%
\end{equation}
as a solutions of Lovelock theory in the Chern-Simons case \cite{MZ}, provided the manifold $d\Sigma_{D-2}^{2}$ fulfils a suitable scalar constraint \cite{DOTTI}, \cite{Dotti:2007az}, \cite{Dotti:2008pp}, \cite{Dotti:2010bw}. It was shown in \cite{COT} that a linear, non-minimally coupled scalar probe, fulfilling reflective boundary conditions has the following spectrum
\begin{equation}
\omega^{2}_n\!=\!\left(  n+\frac{1}{2}+\sqrt{\left(  \frac{D-1}{2}\right)
^{2}\!+m_{\mathrm{eff}}^{2} \ell^{2}}\right)  ^{2}\!\!-\left(  \frac
{D-2}{2}\right)  ^{2}\!\!+Q+\xi\left[  \left(  D-1\right)  \left(  D-2\right)
+\tilde{R}\right]  \,. \label{rrfrec2}%
\end{equation}
where $m_{\mathrm{eff}}^{2}:=m^2 \ell^2-D(D-1)\xi$, $\tilde{R}$ is the Ricci scalar of the Euclidean manifold $\Sigma_{D-2}$ (which we assume constant) and $Q$ stands for an eigenvalue of the Laplace operator on such Euclidean manifold, normalized as $\nabla_\Sigma^2Y=-QY$ ($Q$ being positive if $\Sigma$ is compact and without boundary). Generically, this spectrum will be only asymptotically resonant, nevertheless, in the particular case in which
\begin{equation}
\left(  \frac
{D-2}{2}\right)  ^{2}\!\!-Q-\xi\left[  \left(  D-1\right)  \left(  D-2\right)
+\tilde{R}\right]=0\ ,
\end{equation}
the spectrum will be exactly equispaced. For a given rotational dependence of the scalar probe, i.e. for a fixed value of $Q$, resonance can be achieved for a particular value of the non-minimal coupling parameter $\xi$, providing a new, infinite family of gravitational backgrounds with fully resonant, equispaced spectrum for scalars probes.
\begin{itemize}
\item\textbf{On the universality of the weakly non-linear dynamics in Lovelock theories}
\end{itemize}

Some comments on the backreacting case are in order. For the Einstein-Gauss-Bonnet theory, the scalar field collapse in AdS has been explored \cite{Deppe:2016dcr}. As shown below, remarkably, one can give a general analysis of the perturbative TTF, time averaged approach in a generic Lovelock theory, in AdS. The field
equations of Lovelock theories coupled to a massless scalar (the analysis can
be trivially extended to the massive case) read%

\begin{equation}
\mathcal{E}_{\mu\nu}:=\sum_{k=0}^{[D/2]}\alpha_{k}E_{\mu\nu}^{(k)}-T_{\mu\nu}=0\ ,
\label{resultados-principales:ec-lh}%
\end{equation}
where $T_{\mu\nu}$ stands for the stress-energy tensor of the minimally coupled scalar field $\phi$ and the Lovelock tensor of order $k$ is defined as%

\begin{equation}
E_{\mu\nu}^{(k)}:=-\frac{1}{2^{k+1}}g_{\left(  \mu\right\vert \sigma}\delta_{\left\vert
\nu\right)  \gamma_{1}\cdots \gamma_{2k}}^{\sigma\rho_{1}\cdots \rho_{2k}}R^{\gamma_{1}\gamma_{2}}{}{}%
_{\rho_{1}\rho_{2}}\cdots R^{\gamma_{2k-1}\gamma_{2k}}{}{}_{\rho_{2k-1}A_{2k}}\ .
\label{resultados-principales:tensor-lovelock}%
\end{equation}
Here the couplings $\alpha_{k}$ are dimensionfull. 

Consider a metric of the form%
\begin{equation}
ds_{n}^{2}=g_{\mu\nu}dx^{\mu}dx^{\nu}=g_{ab}^{\left(  2\right)  }\left(
y^c\right)  dy^{a}dy^{b}+F^{2}\left(  y\right)  d\Omega_{S^{n-2}}^{2}\ ,
\end{equation}
where $d\Omega_{S^{n-2}}$ stands for the line element of the $\left(
n-2\right)  $-sphere, $g_{ab}^{\left(  2\right)  }$ is a metric on a
two-dimensional, Lorentzian manifold $M_{2}$, and $F\left(  y\right)  $ is a
scalar on $M_{2}$.

The components of the $p-$th Lovelock tensor along the two-dimensional
manifold $M_{2}$ were explicitly computed in \cite{MAEDA} and read
\begin{align*}
E_{ab}^{\left(  k\right)  }  & =-\frac{k\left(  n-2\right)  !}{\left(
n-2k-1\right)  !}\frac{D_{a}D_{b}F-g_{cd}^{\left(  2\right)  }D^{c}%
FD^{d}F\ g^{(2)}_{ab}}{F}\left(  \frac{1-g_{ef}^{\left(  2\right)  }D^{e}FD^{f}%
F}{F^{2}}\right)  ^{k-1}\\
& -\frac{\left(  n-2\right)  !}{2\left(  n-2k-2\right)  !}g_{ab}^{\left(
2\right)  }\left(  \frac{1-g_{ef}^{\left(  2\right)  }D^{e}FD^{f}F}{F^{2}%
}\right)  ^{k}\ ,
\end{align*}
where $D_{a}$ is the Levi-Civita covariant derivative on $M_{2}$. As usual,
due to diffeomorphism invariance and spherical symmetry the Lovelock equations
along the angles in $S^{n-2}$, are a consequence of the equation along $M_{2}$ and the
equation for the scalar field. The expression for the Lovelock equations on the metric of our interest
\begin{equation}
ds^2=\frac{L^2}{\cos^2(x)}\left[-e^{-2f(t,x)}A(t,x)dt^2+\frac{dx^2}{A(t,x)}+\sin^2(x)d\Omega_{S^{n-2}}^2\right]\ , \label{metriclovelock}
\end{equation}
can be directly obtained by setting $F(y^a)=\tan{x}$ and $g_{ab}^{\left(  2\right)  }$ the metric along the $(t,x)$ directions in \eqref{metriclovelock}. Following \cite{enEGB}, here we will consider the scaled slow time
\begin{equation}
\tau=s_1\varepsilon^{2}t\ ,
\end{equation}
where $s_1$ and $s_2$ (below) are finite constants to be fixed at convenience. We will consider the expansions
\begin{eqnarray}
A(t,\tau,x)&=&1+s_2\varepsilon^2 A_2(t,\tau,x)+\mathcal{O}(\varepsilon^4)\ ,\\
f(t,\tau,x)&=&s_2\varepsilon^2 f_2(t,\tau,x)+\mathcal{O}(\varepsilon^4)\ ,\\
\phi(t,\tau,x)&=&\varepsilon\phi_1(t,\tau,x)+s_2\varepsilon^3\phi_3(t,\tau,x)+\mathcal{O}(\varepsilon^5)\ .
\end{eqnarray}
 At the lowest order, the Lovelock equations determine the AdS radius
$L$ in terms of the couplings $\alpha_{p}$, through the equation%
\begin{equation}
P\left[  L^{-2}\right]  :=\sum_{p=0}^{[D/2]}\frac{\alpha_{p}\left(
-1\right)  ^{p}}{\left(  n-2p-1\right)  !}\left(  \frac{1}{L^{2}}\right)
^{p}=0\ ,\label{poly}%
\end{equation}
which defines the polynomial $P\left[  \xi\right] $. The equation for the scalar field, at the lowest order, determine $\phi_1(t,x)$ as an arbitrary superposition of $n$-dimensional AdS oscilons. At order
$\varepsilon^{2}$ one obtains the following expression for the $\mathcal{E}_{\ t}^{t}$,
$\mathcal{E}_{\ x}^{x}$ and $\mathcal{E}_{\ x}^{t}$ equations in \eqref{resultados-principales:ec-lh}:%
\begin{eqnarray}
-\frac{\left(  n-2\right)!}{2L^2}s_2\frac{dP\left[  L^{-2}\right]  }{d\xi}\left(
\frac{A_{2}^{\prime}}{\tan x}-\frac{\left(  2\cos^{2}x-n+1\right)  }{\sin
^{2}x}A_{2}\right)    & =-\frac{\cos^{2}x}{2L^2}\left(  \dot{\phi}_{1}^{2}+\phi^{\prime 2}
_{1}\right)  \ , \label{ttflovelock1}\\
-\frac{\left(  n-2\right)!}{2L^2}s_2\frac{dP\left[  L^{-2}\right]  }{d\xi
}\frac{\dot{A}_2}{\tan x}  & =-\frac{\cos^{2}x}{L^2}\phi_{1}	^{\prime}\dot{\phi}_{1}\ ,\label{ttflovelock2}\\
-\frac{\left(  n-2\right)  !}{2L^2}s_2\frac{dP\left[  L^{-2}\right]  }{d\xi}\left(
\frac{A_{2}^{\prime}}{\tan x}-\frac{\left(  2\cos^{2}x-n+1\right)  }{\sin
^{2}x}A_{2}-\frac{2f_{2}^{\prime}}{\tan x}\right)    & =\frac{\cos^{2}x}{2L^2}\left(
\dot{\phi}_{1}^{2}+\phi^{\prime 2}
_{1}\right)  \ .\label{ttflovelock3}
\end{eqnarray}
where (') and ($\cdot$) denote derivative with respect to $x$ and $t$, respectively.  No derivatives with respect to the slow time appear at this order. From these equations one can solve $A_2(t,x)$ and $f_2(t,x)$ as in GR. Then, substituting this at the next non-trivial order in the Klein-Gordon equation, one obtains \begin{equation}
\ddot{\phi}_{3}+\mathcal{L}\left[  \phi_{3}\right]  +\frac{2s_{1}}{s_{2}%
}\partial_{t}\partial_{\tau}\phi_{1}=\phi_{1}^{\prime}\dot{A}_{2}+\phi
_{1}^{\prime}A_{2}^{\prime}-\dot{\phi}_{1}\dot{f}_{2}-\phi_{1}^{\prime}%
f_{2}^{\prime}+2\left(  f_{2}-A_{2}\right)  \mathcal{L}\left[  \phi
_{1}\right]\label{paraphi3}
\end{equation}
where the action of the operator $\mathcal{L}\left[  \phi_{i}\right]  $ is
defined as%
\begin{equation}
\mathcal{L}\left[  \phi_{i}\right]  :=-\frac{\partial^{2}\phi_{i}}{\partial
x^{2}}-\frac{\left(  n-2\right)  }{\cos x\sin x}\frac{\partial\phi_{i}%
}{\partial x}\ .
\end{equation}

Therefore, as in reference \cite{enEGB} for the Einstein-Gauss-Bonnet case in five dimensions, but now in the whole family of Lovelock theories, setting\begin{equation}
s_1=s_2=\left(\frac{dP\left[  L^{-2}\right]  }{d\xi}\right)^{-1}
\end{equation}
we see that the equations for the TTF approach, for a generic Lovelock theory \eqref{ttflovelock1}-\eqref{paraphi3} take
exactly the same functional form than the equations in GR, provided we are
expanding about AdS with a curvature corresponding to a simple zero of the
polynomial (\ref{poly}), for which $\frac{dP\left[  L^{-2}\right]  }{d\xi}$ is non-vanishing. It is interesting to see that this polynomial,
completely controls the perturbative dynamics in Lovelock theories. On the other hand, for the wormhole studied in this work, the asymptotic AdS curvature radius exactly cancels the derivative of the polynomial (this occurs for any Chern-Simons theory within the Lovelock family) and therefore, the
perturbative approach, in the backreacting situation does not apply. This is why we have focused on the probe limit of the scalar. Notwithstanding, the equations presented above for Lovelock theories are a signal of the
universality of the weakly-nonlinear dynamics up to times of order $\varepsilon^{-2}$, captured by the TTF, for generic values of the
couplings, when the higher curvature curvature terms belong to the
Lovelock family. This was shown in \cite{enEGB} (see also \cite{Buchel:2015lla}) for the Einstein-Gauss-Bonnet case, and here we have shown that such results extend to the whole family of Lovelock theories for generic values of the couplings. These results extend also for the family of Quasitopological Gravities \cite{QTGOriginal}, \cite{Myers:2010ru}, \cite{Dehghani:2011vu}, \cite{Cisterna:2017umf}, since the dynamics on the spherically symmetric dynamical scenarios with matter, can be obtained from the same formulae \eqref{ttflovelock1}-\eqref{ttflovelock3}, by including extra terms in the polynomial. All the mentioned theories admit Birkhoff's theorems, therefore on a spherically symmetric scenario, the dynamics is completely driven by the scalar.

It is well-known that for Lovelock theories containing a $k$-th order term in dimension $n=2k+1$, the maximally symmetric AdS vacuum is gapped with respect to the smallest black hole (see e.g. \cite{Garraffo:2008hu}, \cite{Charmousis:2008kc}), as it occurs for the BTZ  black hole \cite{BTZ}, a feature that is captures by the numerical evolutions (see e.g. \cite{Bizon:2013xha}, \cite{Pretorius:2000yu} for the $2+1$ case and \cite{Deppe:2012wk} for $4+1$ dimensions). It is interesting to note that the structure of the TTF dynamics, at times $\varepsilon^{-2}$, being universal for Lovelock theories with generic couplings, does not capture this gap\footnote{In \cite{Menon:2015oda} it was noted that in a direct perturbative at order $\varepsilon^{4}$ the presence of a Gauss-Bonnet term cannot be scaled out.}.   

Finally, it would be interesting to explore the dynamics of a scalar
 collapse to a black hole in other asymptotically AdS solitons with two ends,
including
 backreaction. The recently constructed analytic, wormhole solution
 of General Relativity with a negative cosmological constant \cite{Anabalon:2018rzq}
 defines a perfect scenario to initiate such
 exploration,
 since the Einstein-Klein-Gordon system leads to a well-posed initial boundary 
value problem in asymptotically AdS spacetimes.

\section{Acknowledgements}

We would like to thank F. Correa, O. Evnin, O. Fierro, S. Green, L. Lehner and H. Maeda for enlightening comments.  This work was partially funded by FONDECYT Grants 1150246 and 1170279.

\end{document}